\batchmode
\makeatletter
\def\input@path{{D:/DATA/TiSe2/TiSe2_PRL_reply_2/TiSe2_PRL_DefinitiveEdition/}}
\makeatother
\documentclass[twocolumn,prl,showpacs,superscriptaddress,longbibliography]{revtex4-2}
\usepackage[latin9]{inputenc}
\usepackage{color}
\usepackage{amsmath}
\usepackage{graphicx}
\usepackage[pdftex,unicode=true,pdfusetitle,
 bookmarks=true,bookmarksnumbered=false,bookmarksopen=false,
 breaklinks=false,pdfborder={0 0 0},pdfborderstyle={},backref=false,colorlinks=true]
 {hyperref}


\begin{document}
\title{Dramatic Plasmon Response to the Charge-Density-Wave Gap Development
in 1\emph{T-}TiSe$_{2}$}
\author{Zijian Lin}
\thanks{Equally contributed to this work.}
\affiliation{Beijing National Laboratory for Condensed Matter Physics and Institute
of Physics, Chinese Academy of Sciences, Beijing 100190, China}
\affiliation{School of Physical Sciences, University of Chinese Academy of Sciences,
Beijing 100049, China}
\author{Cuixiang Wang}
\thanks{Equally contributed to this work.}
\affiliation{Beijing National Laboratory for Condensed Matter Physics and Institute
of Physics, Chinese Academy of Sciences, Beijing 100190, China}
\affiliation{School of Physical Sciences, University of Chinese Academy of Sciences,
Beijing 100049, China}
\author{A.~Balassis}
\affiliation{Department of Physics and Engineering Physics, Fordham University,
441 East Fordham Road, Bronx, NY 10458, USA}
\author{J.\,P.~Echeverry}
\affiliation{Universidad de Ibagué, Carrera 22 Calle 67 B, Av. Ambalá, Ibagué-Tolima,
Colombia}
\author{A.\,S.~Vasenko}
\affiliation{HSE University, 101000 Moscow, Russia}
\affiliation{I. E. Tamm Department of Theoretical Physics, P. N. Lebedev Physical
Institute, Russian Academy of Sciences, 119991 Moscow, Russia}
\author{V.\,M.~Silkin}
\affiliation{Donostia International Physics Center (DIPC), 20018 San Sebastián/Donostia,
Basque Country, Spain}
\affiliation{Departamento de Polímeros y Materiales Avanzados: Física, Química
y Tecnología, Facultad de Ciencias Químicas, Universidad del País
Vasco UPV/EHU, Apartado 1072, 20080 San Sebastián/Donostia, Basque
Country, Spain}
\affiliation{IKERBASQUE, Basque Foundation for Science, 48013 Bilbao, Basque Country,
Spain}
\author{E.\,V.~Chulkov}
\affiliation{Donostia International Physics Center (DIPC), 20018 San Sebastián/Donostia,
Basque Country, Spain}
\affiliation{Departamento de Polímeros y Materiales Avanzados: Física, Química
y Tecnología, Facultad de Ciencias Químicas, Universidad del País
Vasco UPV/EHU, Apartado 1072, 20080 San Sebastián/Donostia, Basque
Country, Spain}
\affiliation{HSE University, 101000 Moscow, Russia}
\author{Youguo Shi}
\affiliation{Beijing National Laboratory for Condensed Matter Physics and Institute
of Physics, Chinese Academy of Sciences, Beijing 100190, China}
\affiliation{Songshan Lake Materials Laboratory, Dongguan, Guangdong 523808, China}
\author{Jiandi Zhang}
\affiliation{Beijing National Laboratory for Condensed Matter Physics and Institute
of Physics, Chinese Academy of Sciences, Beijing 100190, China}
\author{Jiandong Guo}
\email{jdguo@iphy.ac.cn}

\affiliation{Beijing National Laboratory for Condensed Matter Physics and Institute
of Physics, Chinese Academy of Sciences, Beijing 100190, China}
\affiliation{School of Physical Sciences, University of Chinese Academy of Sciences,
Beijing 100049, China}
\affiliation{Songshan Lake Materials Laboratory, Dongguan, Guangdong 523808, China}
\author{Xuetao Zhu}
\email{xtzhu@iphy.ac.cn}

\affiliation{Beijing National Laboratory for Condensed Matter Physics and Institute
of Physics, Chinese Academy of Sciences, Beijing 100190, China}
\affiliation{School of Physical Sciences, University of Chinese Academy of Sciences,
Beijing 100049, China}
\affiliation{Songshan Lake Materials Laboratory, Dongguan, Guangdong 523808, China}
\begin{abstract}
1\emph{T}-TiSe$_{2}$ is one of the most studied charge density wave
(CDW) systems, not only because of its peculiar properties related
to the CDW transition, but also due to its status as a promising candidate
of exciton insulator signaled by the proposed plasmon softening at
the CDW wave vector. Using high-resolution electron energy loss spectroscopy,
we report a systematic study of the temperature-dependent plasmon
behaviors of 1\emph{T}-TiSe$_{2}$. We unambiguously resolve the plasmon
from phonon modes, revealing the existence of Landau damping to the
plasmon at finite momentums, which does not support the plasmon softening
picture for exciton condensation. Moreover, we discover that the plasmon
lifetime at zero momentum responds dramatically to the bandgap evolution
associated with the CDW transition. The interband transitions near
the Fermi energy in the normal phase is demonstrated serving as a
strong damping channel of plasmons, while such a channel in the CDW
phase is suppressed due to the CDW gap opening, which results in the
dramatic tunability of the plasmon in semimetals or small-gap semiconductors.
\end{abstract}
\maketitle
In a charge density wave (CDW) material, the CDW gap development,
which is often served as the order parameter to characterize the CDW
transition \citep{Gruner1988,Zhu2015a}, can strongly influence the
emergent phenomena of the system. 1\emph{T-}TiSe$_{2}$, a quasi-two-dimensional
layered material, undergoes a three-dimensional second order CDW transition
at $T_{\text{c}}\sim200$ K with $\textbf{\emph{\textbf{q}}}_{\text{CDW}}=(1/2,0,1/2)$
\citep{Salvo1976}. The CDW origin of 1\emph{T-}TiSe$_{2}$ was described
by several different mechanisms \citep{Salvo1976,Hughes1977,Wilson1978a}.
One of the most-studied scenarios is the electron-phonon coupling
\citep{Hughes1977,WAKABAYASHI1978,Holt2001,Kidd2002,Rossnagel2002,Weber2011,Hildebrand2014,Porer2014,Hildebrand2016},
which is the CDW origin in many quasi-two-dimensional systems \citep{Zhu2015a}.
On the other hand, the CDW order in 1\emph{T-}TiSe$_{2}$ is proposed
to be induced by the formation of the exciton insulator (EI) \citep{Wilson1978,Pillo2000,Cercellier2007,Qian2007,Monney2009,Monney2010,Vorobeva2011,Kogar2017}.
And some recent studies claimed that the EI and the electron-phonon
coupling may cooperatively induce the CDW transition \citep{Wezel2010,Wezel20101,Wezel2011,Monney2011,May2011,Monney2016}.

\begin{figure}
\includegraphics[width=8.6cm]{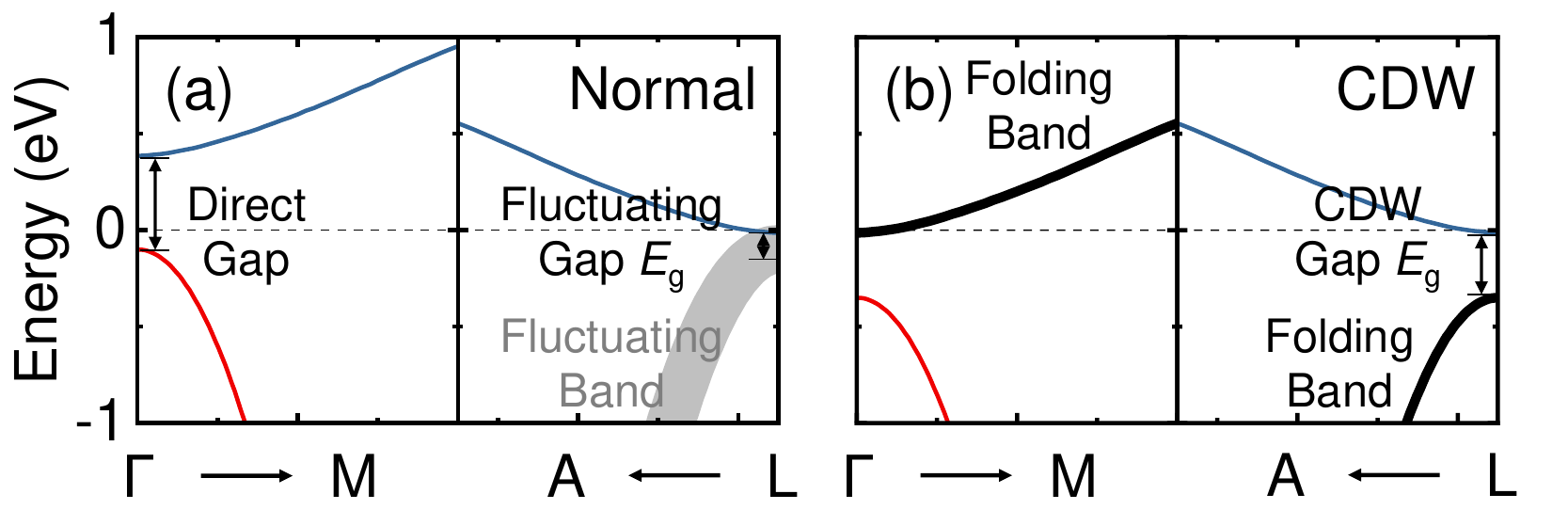}\caption{\label{5}Schematics of the band structure in 1\emph{T-}TiSe$_{2}$.
(a) band structure in the normal phase. The red and blue lines represent
the valence and conduction bands, respectively. The fluctuating band
is represented by the gray shadow. The dashed lines indicate the Fermi
level. (b) band structure in the CDW phase. Black lines represent
the folding bands due to the distortion.}
\end{figure}

Understanding the the band structure evolution is essential for identifying
the origin of the CDW transition \citep{Monney2015,Bok2021}. In the
normal phase above $T_{\text{c}}$ {[}Fig. \ref{5}(a){]}, 1\emph{T-}TiSe$_{2}$
is a semiconductor with a small indirect gap \citep{Monney2015} (or
a semimetal with a small band overlap \citep{Rossnagel2002}), where
the bottom of the conduction band is almost tangent to the Fermi level
$E_{\text{F}}$ \citep{Bianco2015,Wegner2020}. Interestingly, there
exists a CDW fluctuating band above $T_{\text{c}}$ \citep{Wilson1978,Holt2001,Kidd2002,Monney2012,Chen2016},
which may provide a precursor of the CDW gap \citep{Miyahara1995}.
In the CDW phase below $T_{\text{c}}$ {[}Fig. \ref{5}(b){]}, the
gap opens by the valence band gradually shifting toward to higher
binding energy, leaving the conduction band still tangent to $E_{\text{F}}$
and accompanying with a CDW band folding \citep{Pillo2000,Kidd2002,Rossnagel2002,Cercellier2007,Monney2012}.
Due to the coexistence of the metallic nature and CDW gap in both
the normal and CDW phases, the plasmon provides a good window to visualizing
the effect of the CDW gap development on the electronic properties.
Recently, the plasmon softening at $\textbf{\emph{\textbf{q}}}_{\text{CDW}}$
around $T_{\text{c}}$ was proposed to serve as the signature of the
EI \citep{Kogar2017}. However, a recent theoretical study attributed
the seemingly plasmon softening signal to the interband transitions
\citep{Lian2019}.

In this letter, using the momentum-resolved high resolution electron
energy loss spectroscopy (HREELS) with the capability of two-dimensional
energy-momentum mapping \citep{Zhu2015}, we systematically measured
the plasmon behaviors in 1\emph{T-}TiSe$_{2}$. Our results unambiguously
resolve the plasmon from phonon modes and demonstrate the existence
of Landau damping \citep{Haque1973,Pandey1974,raether2006excitation}
to the plasmon at finite momentums within the full temperature range,
revealing that there is no plasmon softening at $\textbf{\emph{\textbf{q}}}_{\text{CDW}}$.
Extraordinarily, we discover that the plasmon at zero momentum responds
to the CDW transition dramatically, from a broad feature covering
the energy of $50\sim150$ meV to a sharp feature with the well-defined
resonant frequency at $\sim50$ meV in the low-temperature CDW phase.
Such a wide-range tunability is attributed to the gap opening associated
with the CDW transition that suppresses the interband damping channels
of the plasmon.

\begin{figure*}
\centering{}\includegraphics[width=1\textwidth]{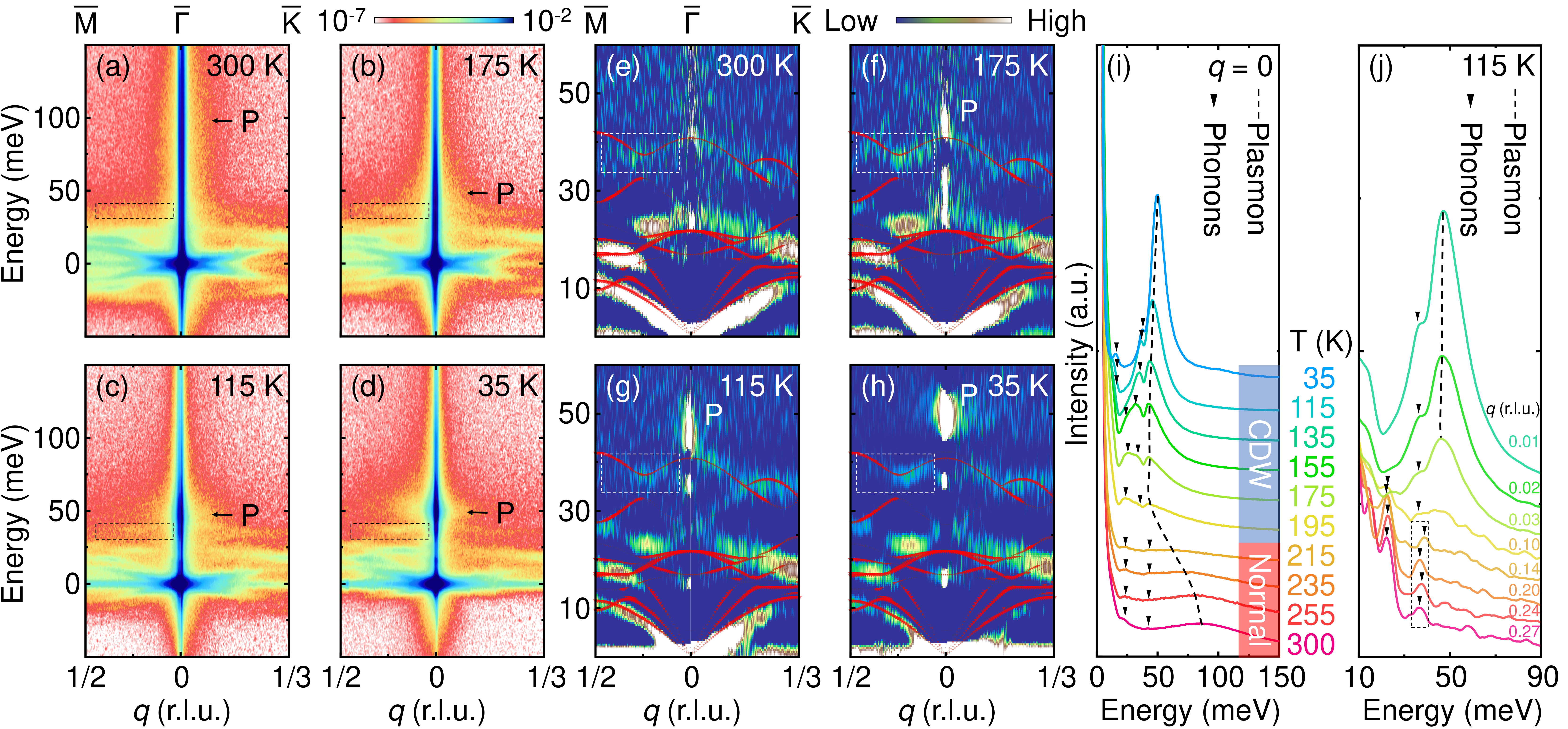}\caption{\label{1}Temperature (\emph{T})-dependent HREELS results in 1\emph{T-}TiSe$_{2}$.
(a)-(d) \emph{E}-$\textbf{\emph{\textbf{q}}}_{//}$ mappings of HREELS
at 300, 175, 115, and 35 K, respectively. At the ${\rm \overline{\Gamma}}$
point, the plasmon is labeled by P. Dashed-line rectangles show the
typical unchanged phonon dispersions close to the plasmon. The momenta
are represented in the form of (\emph{q}, 0) and (\emph{q}, \emph{q})
along ${\rm \overline{\Gamma}}$-${\rm \overline{M}}$ and ${\rm \overline{\Gamma}}$-${\rm \overline{K}}$
directions with the reciprocal lattice unit (r.l.u.) as the unit.
The color scale corresponds to the logarithmic intensity. The positive
and negative energy ranges represent the Stokes and anti-Stokes scatterings,
respectively. (e)-(h) Second differential images of (a)-(d), respectively.
The red triangles are the surface phonon dispersions calculated by
the slab method. The size of triangles indicates the spectral weight
of the surface amplitude of phonons. White dashed-line rectangles
are the same as the dashed-line rectangles in (a)-(d). (i) Stack of
the EDCs at the ${\rm \overline{\Gamma}}$ point at different temperatures.
The triangles and dashed line are the guides to eyes of phonons and
the plasmon, respectively. (j) Stack of the EDCs at 115 K at different
\emph{q}'s. The intensities of EDCs at \emph{q} = 0.01, 0.02, and
0.03 r.l.u. is multiplied by 0.03, 0.08, and 0.2 for better presentation,
respectively. The dashed-line rectangle is the same as the dashed-line
rectangles in (a)-(d).}
\end{figure*}

\emph{Temperature-dependent HREELS results.} - The measurements were
performed on cleaved 1\emph{T-}TiSe$_{2}$ single-crystalline samples
\emph{in situ} in a HREELS system with reflected scattering geometry.
The detailed experimental methods and sample characterizations are
described in the Supplementary Material (SM) \citep{SM}. The HREELS
data were collected with the sample temperature varying from 35 to
300 K along the two high symmetry directions $\overline{\Gamma}$-${\rm \overline{M}}$
and $\overline{\Gamma}$-${\rm \overline{K}}$ in the surface Brillouin
zone (BZ). The incident electron beam energy $E_{\text{i}}$ from
7 to 110 eV (110 eV data presented in the main paper, while others
in the SM \citep{SM}) is used , with a typical energy resolution
of 3 meV. Figure \ref{1} shows the HREELS results of 1\emph{T-}TiSe$_{2}$
at various temperatures. The temperature-dependent \emph{E}-\emph{q}$_{//}$
mappings along $\overline{\Gamma}$-${\rm \overline{M}}$ and $\overline{\Gamma}$-${\rm \overline{K}}$
directions are presented in Figs. \ref{1}(a)-(d).

In HREELS, the strong intensity distributions near the $\overline{\Gamma}$
point are dominated by the dipole scattering, and the relatively weak
features away from the $\overline{\Gamma}$ point to the BZ boundary
correspond to the impact scattering regime \citep{ibach2013electron}.
Then, it is obvious that there are two kinds of distinct energy loss
features divided by the energy of 45 meV at 300 K. The loss features
higher than 45 meV, labeled as P, are only located near the $\overline{\Gamma}$
point, which are pure dipole scattering features and regarded as the
plasmon originating from the charge carrier both in the normal and
CDW phases \citep{Liang1979,Li2007}. The loss features lower than
45 meV exist throughout the BZ, which are typical impact scattering
signals from phonons \citep{ibach2013electron}. To show their dispersions
more clearly, Figs. \ref{1}(e)-(h) display the second differential
images of the original spectra superimposed to the calculated surface
phonon dispersions (red lines) using an 11-layer slab model (see the
details in the SM \citep{SM}). Overall, the calculated phonon dispersions,
especially the optical phonon branches, match well with the experimental
results. 

The evolution of the plasmon with temperature is illustrated by a
stacking plot of the loss features at the $\overline{\Gamma}$ point
{[}Fig. \ref{1}(i){]}. As the temperature decreases from 300 K to
$T_{\text{c}}$, the energy of the plasmon gets closer to the optical
phonon. When the temperature further decreases below $T_{\text{c}}$,
the energy of the plasmon slightly increases and the linewidth of
the plasmon changes dramatically. These phenomena will be discussed
later in more detail.

Compared to the previous HREELS study in Ref. \citep{Kogar2017},
our results show similar temperature-dependent plasmon behavior at
the $\overline{\Gamma}$ point, but surprisingly, we did not observe
the plasmon softening at $\textbf{\emph{\textbf{q}}}_{\text{CDW}}$.
In our energy-momentum mappings at any temperature, the plasmon only
exists near the $\overline{\Gamma}$ point and does not disperse to
the low-energy range at the BZ boundary. In detail, Fig. \ref{1}(j)
shows dispersion behaviors of the plasmon and phonons in the \emph{q}
space along the $\overline{\Gamma}$-${\rm \overline{M}}$ direction
\footnote{In the CDW phase, the plasmon line shape is modulated by phonons above
115 K, since the plasmon energy is close to phonons.}. The plasmon decays from a sharp peak at \emph{q} = 0.01 r.l.u. to
a weaker peak at \emph{q} = 0.03 r.l.u. As \emph{q} increases, the
plasmon is not a well-defined peak anymore at $q=0.10$ r.l.u. and
disappears (indistinguishable from noise) beyond \emph{q} = 0.14 r.l.u.
The reported energy loss signal over the entire BZ at 17 K in Ref.
\citep{Kogar2017}, interpreted as a dispersionless plasmon, could
be from diffusion scattering \citep{ibach2013electron}. In our results,
the delicate optical phonon dispersions {[}those highlighted by the
dashed-line rectangles in Fig. \ref{1}{]}, which were not observed
in Ref. \citep{Kogar2017}, can be clearly resolved from the plasmon.
The energy range of these optical phonon branches are close to the
claimed softening plasmon energy, so the softening plasmon dispersion
around $T_{\text{c}}$ \citep{Kogar2017} seems from the envelope
of phonon signals. Details of the comparison are described in the
SM \citep{SM}.

\begin{figure}
\noindent \begin{raggedright}
\includegraphics[width=8.6cm]{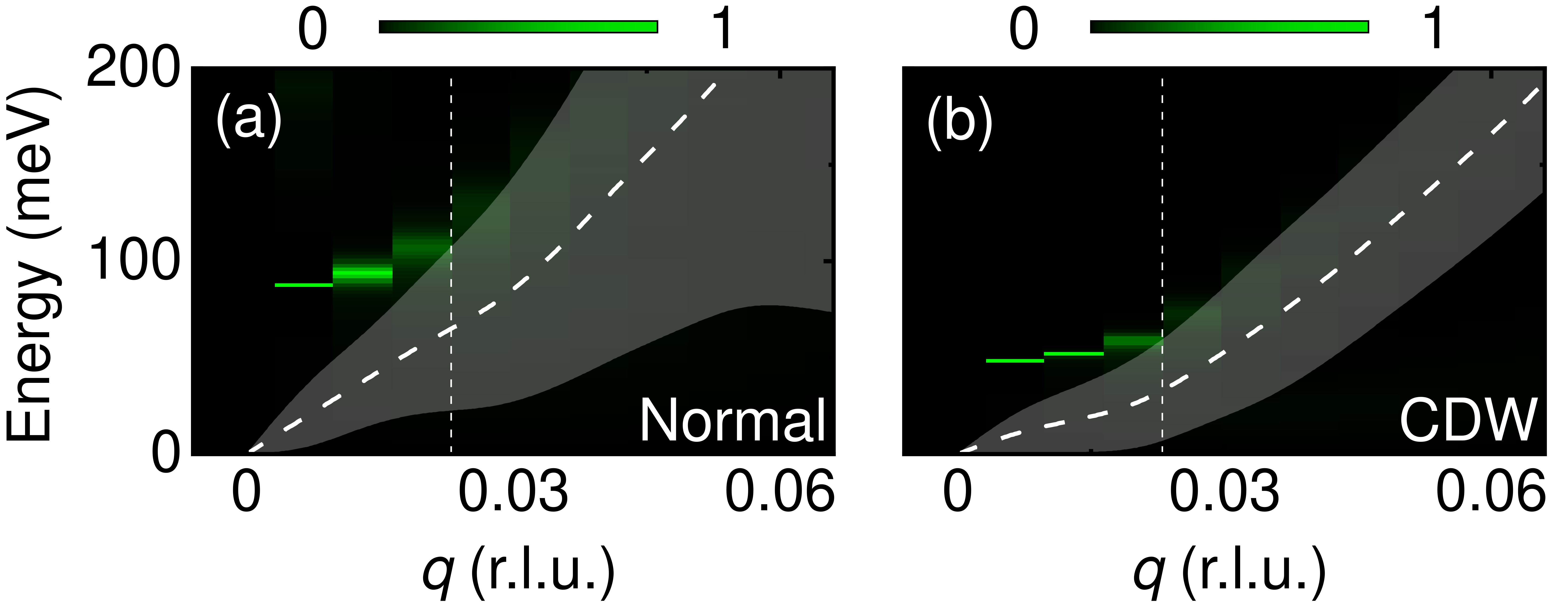}
\par\end{raggedright}
\caption{\label{2} Theoretical SPE in 1\emph{T-}TiSe$_{2}$. (a) and (b) Calculated
loss function mappings of the normal and CDW state, respectively.
The gray shadowed areas centered at thick white dashed lines are the
calculated SPE regions. The green color mappings in loss functions
are plasmons. Thin dashed lines highlight calculated critical momenta
$q_{\text{c}}$.}
\end{figure}

\emph{Landau damping of the plasmon.} - The observed plasmon damping
behaviors can be well understood by the calculations of the loss functions,
which were carried in the framework of time-dependent density functional
theory (see details in the SM \citep{SM}). For simplicity, the calculated
band structure in the CDW phase does not include the effects of band
folding and renormalization. The calculated single particle excitation
(SPE) regions for the normal and CDW band structures are indicated
by white dashed line and shadow areas centered at thick white dashed
lines in Fig. \ref{2}, and plasmon dispersions are represented by
color mappings of the calculated loss functions. At the lowest \emph{q},
plasmons are almost delta functions. However, plasmons broaden and
weaken rapidly approaching SPE borders. Beyond the calculated $q_{\text{c}}=0.024$
r.l.u. \footnote{The $q_{c}$ is defined by the minimum of the momentum where there
is no zero point in the real part of the calculated dielectric function.
See details in the Supplementary Materials \citep{SM}}, plasmons are not well-defined collective excitations anymore due
to the strong decay to SPEs. This explains the experimentally observed
fast decay of the plasmon at small momentum.

\emph{Damping of the plasmon at the long wavelength limit.} - Notice
there is dramatic change of plasmon behaviors across $T_{\text{c}}$
at the long wavelength limit (near $q=0$) , especially the linewidth
{[}Fig. \ref{1}(i){]}. In the normal state, the relatively broad
linewidth indicates the strong plasmon damping above $T_{\text{c}}$.
In contrast, in the CDW state, the linewidth is much smaller, demonstrating
that the plasmon damping is largely suppressed below $T_{\text{c}}$.
Moreover, the calculated loss function near $q=0$ is almost a delta
function without any damping in Fig. \ref{2} in both the normal and
CDW phases, suggesting that an analysis beyond the framework used
above (which excludes the CDW band folding) is necessary.

To characterize the temperature-dependent damping behavior at $q=0$,
we plotted the plasmon lifetime parameter $\text{\ensuremath{\text{\ensuremath{\tau}}_{\text{pl}}=(\Delta E_{\text{pl}}/E_{\text{pl}})^{-1}}}$
in Fig. \ref{3}(a), where $E_{\text{pl}}$ and $\Delta E_{\text{pl}}$
are the energy and linewidth of the plasmon, respectively. The damping
of the plasmon is slightly enhanced when the temperature decreases
from room temperature toward $T_{\text{c}}$, and as the temperature
drops further below $T_{\text{c}}$, the damping is substantially
suppressed, and even becomes much weaker than that at 300 K. Meanwhile,
we noticed that the lifetime of the plasmon and the CDW gap as the
order parameter (adopted from the resonant inelastic X-ray scattering
results \citep{Monney2012a}) have a synchronous temperature-dependent
behavior, implying that the opening of the CDW gap is connected to
the enhancement of the plasmon lifetime, that is the depression of
the plasmon damping below $T_{\text{c}}$.

In classical dielectric theory, the effect of the gap (or interband
transitions) on plasmons is approximated by the high-frequency dielectric
constant $\text{\ensuremath{\epsilon_{\infty}}}$, and the plasmon
energy or frequency is given by $E_{\text{pl}}=\hbar\omega_{\text{pl}}=\hbar(\frac{ne^{2}}{m^{*}\ensuremath{\epsilon_{\infty}}})^{1/2}$,
where $n$, $e$, and $m^{*}$ are the carrier concentration, the
charge of an electron, and the effective electron mass, respectively.
In Fig. \ref{3}(b), we compared the fitted plasmon energy at $q=0$
from our experiment and the carrier concentration in Ref. \citep{Knowles2020}.
Although both $E_{\text{pl}}$ and $n^{1/2}$ as functions of temperature
show abrupt changes at $T_{\text{c}}$, the discrepancy between them
clearly indicates that, besides the carrier density, there must be
other factors that are responsible for the evolution of the plasmon
energy. We divided $(E_{\text{pl}}^{2})^{-1}$ by $n^{-1}$ to exclude
the effect of the carrier concentration, and obtained the temperature
dependence of the multiplication of $\epsilon_{\infty}$ and $m^{*}$
{[}Fig. \ref{3}(c){]}. The previous experiment \citep{Monney20101}
did not show an observable change of $m^{*}$ near $T_{\text{c}}$,
so the additional drop in plasmon energy compared to $n^{1/2}$ on
cooling across $T_{\text{c}}$ should be attributed to an abrupt increase
in $\text{\ensuremath{\epsilon_{\infty}}}$ related to the CDW gap.
Besides, the plasmon dispersion gradually flattens upon cooling, from
a conventional parabolic shape in the normal phase to a dispersionless
flat line in the CDW phase, as shown in Fig. \ref{3}(d).

\emph{Evolution of the plasmon with the CDW gap.} - To explain the
temperature-dependent plasmon features, we show a phenomenological
model where the plasmon evolves with the development of the CDW gap.
In this model, we employed a simple dielectric theory to simulate
the influence of the interband transition across the CDW gap to the
dispersion and damping of the plasmon. The dielectric function can
be expressed as follows (see details in the SM \citep{SM}):
\begin{figure}
\noindent \begin{raggedright}
\includegraphics[width=8.6cm]{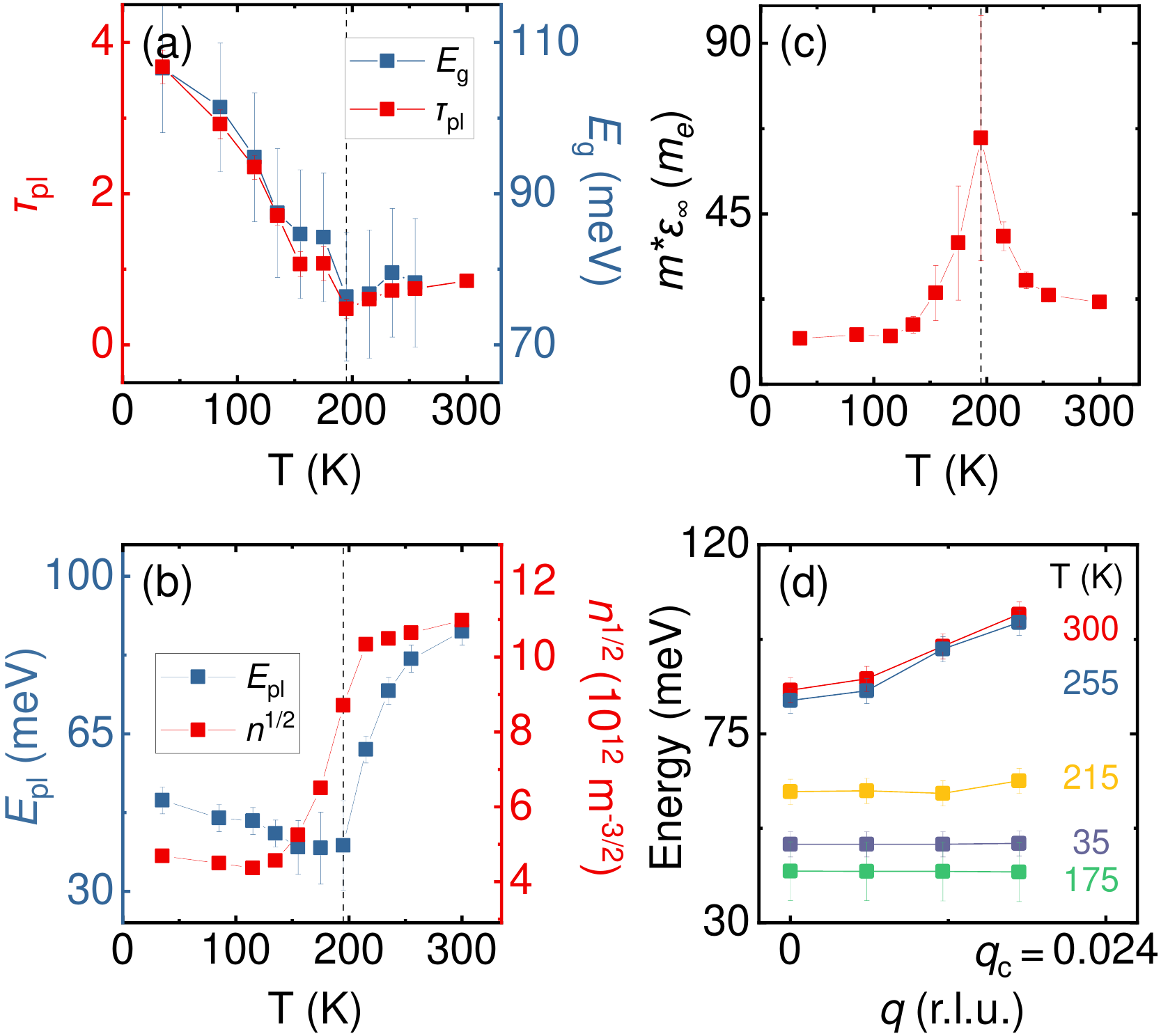}
\par\end{raggedright}
\caption{\label{3}\emph{Temperature-dependent plasmon features} in 1\emph{T-}TiSe$_{2}$.
(a) Temperature dependence of the dimensionless lifetime parameter
$\tau_{\text{pl}}$ of the plasmon and the CDW gap $E_{\text{g}}$
in Ref. \citep{Monney2012a}. The dashed line highlights the $T_{\text{c}}$.
(b) Temperature dependence of the plasmon energy $E_{\text{pl}}$
and the square root of the carrier concentration $n^{1/2}$ in Ref.
\citep{Knowles2020}, respectively. (c) Temperature dependence of
the multiplication of the effective carrier mass $m^{*}$ and the
high frequency dielectric constant $\epsilon_{\infty}.$ (d) Temperature
dependence of plasmon dispersion below $q_{\text{c}}$.}
\end{figure}

\begin{align*}
\epsilon(q,\omega)= & \ensuremath{1-\frac{\omega_{\text{pl}}^{2}(q)}{\omega^{2}}}+\frac{\omega_{\text{intra}}^{2}(q)}{\omega^{2}-\omega_{\text{intra}}^{2}(q)+\text{i}\Gamma_{\text{intra}}\omega}\\
 & +\frac{\omega_{\text{pl}}^{2}(q)}{\omega^{2}-\omega_{\textrm{CDW}}^{2}(q)+\text{i}\Gamma_{\text{CDW}}\omega}.
\end{align*}

\begin{figure*}
\noindent \begin{centering}
\includegraphics[width=17.2cm]{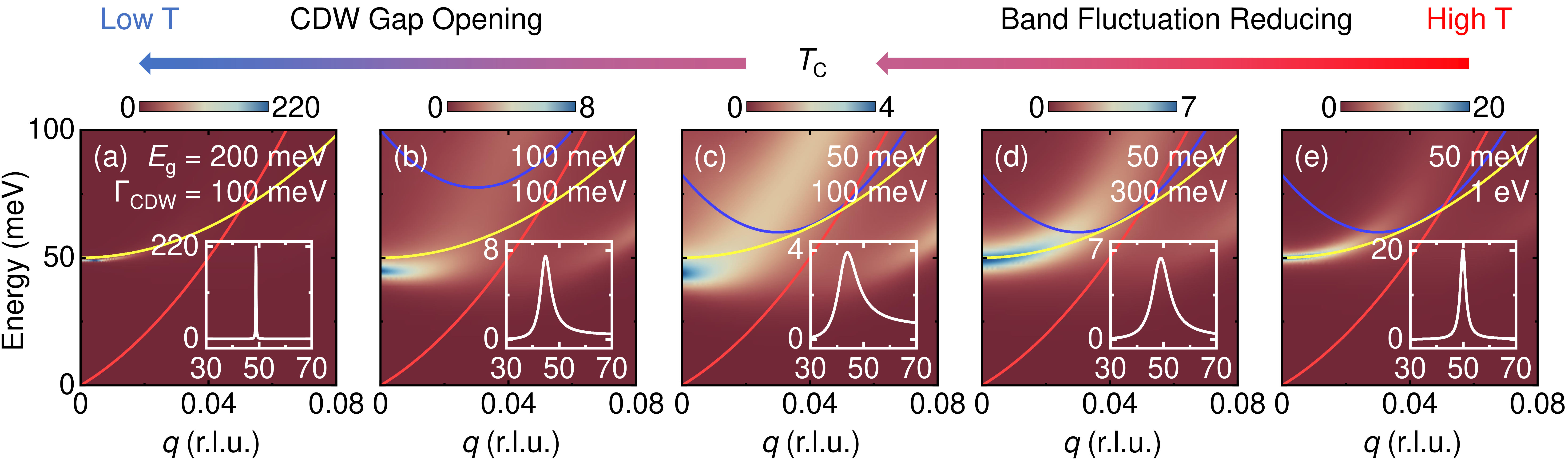}
\par\end{centering}
\caption{\label{4}Loss functions simulated by a Drude-Lorentz model. (a)-(e)
Loss functions $\text{Im}[-\epsilon(q,E)^{-1}]$ with different CDW
gaps $E_{\text{g}}=200$, 100, 50, 50, and 50 meV and interband damping
parameters $\Gamma_{\text{CDW}}=100$, 100, 100, 300 meV, and 1 eV,
respectively. The yellow, red, and blue lines highlight the input
parameters: energies of a partly screened plasmon, the intraband transitions,
and the interband transitions across the CDW gap, respectively. Inset:
the line shape of the loss function at \emph{q} = 0. The labels on
the horizontal and vertical axes are Energy (meV) and Loss Function,
respectively.}
\end{figure*}

The first two terms $(1-\frac{\omega_{\text{pl}}^{2}(q)}{\omega^{2}})$
represents a partly screened plasmon by the higher-energy interband
transitions other than ones across the CDW gap. In this simulation,
the plasmon dispersion $\omega_{\text{pl}}(q)$ follows a parabolic
form (yellow lines in Fig. \ref{4}) in the random phase approximation.
The third term represents the intraband transitions {[}i.e. the Landau
damping in Fig. \ref{2}{]}, where $\omega_{\text{intra}}$ corresponds
to the SPE boundary \footnote{In this simulation, the interband and intraband transition refer to
the SPE boundary, but the SPE continuum in the region of large \emph{q}
is not considered.}, and $\Gamma_{\text{intra}}$ is the the linewidth of the intraband
transitions. The above parameters in the plasmon and intraband terms
are kept constant in this simulation. The last term represents the
interband transitions purely originating from the CDW gap (blue lines
in Fig. \ref{4}), where $\hbar\omega_{\textrm{CDW}}(q)=E_{\text{g}}+(E_{\text{F}}-E_{\text{cb}})+C(q-k_{\text{F}})^{2}$
with $E_{\text{g}}$ the size of the gap, $E_{\text{cb}}$ the energy
of the conduction band bottom, and \emph{C }the constant related to
the band slope (see details in the SM \citep{SM}). $\Gamma_{\text{CDW}}$
is the linewidth of the interband transition across the CDW gap. The
loss functions $\text{Im}[-\epsilon(q,\omega)^{-1}]$ are plotted
in Fig. \ref{4} with different CDW gap parameters.

In the normal phase, the lowest interband transition is contributed
by the hopping from the CDW fluctuating band \citep{Wilson1978,Miyahara1995,Holt2001,Kidd2002,Monney2012,Chen2016}
to the conduction band at $E_{\text{F}}$ {[}shown as Fig. \ref{5}(a){]}.
Referring to the resonant inelastic X-ray scattering results \citep{Monney2012a},
we let $E_{\text{g}}=50$ meV. At high temperature {[}Fig. \ref{4}(e){]},
the smeared fluctuating band is corresponding to a large $\hbar\Gamma_{\textrm{CDW}}=1$
eV \footnote{The real $\Gamma_{\text{CDW}}$ may be not so large. To emphasize
the evolution of the plasmon, we chose such an extreme value.} as a weak damping channel to the plasmon. As the temperature decreases,
the band fluctuation becomes smaller, i.e., the fluctuating band tends
to be the folding band as a stronger damping channel to the plasmon,
which corresponds to the smaller $\Gamma_{\text{CDW}}$ {[}Figs. \ref{4}(c)-(e){]}.
In this process, the plasmon energy is getting slightly lower, the
height is suppressed, and the linewidth broadens. Close to $T_{\textrm{c}}$,
with $\hbar\Gamma_{\textrm{CDW}}=100$ meV, the peak of the plasmon
becomes asymmetric {[}Fig. \ref{4}(c){]}, indicating the interference
with the interband transition across the CDW gap.

In the CDW phase, the CDW fluctuation changes into a well-defined
back-folding band {[}shown as Fig. \ref{5}(b){]}. As the temperature
decreases, the $E_{\text{g}}$ increases \citep{Monney2012a,Chen2016}
and the interband transitions move away from the plasmon thus reducing
the corresponding plasmon damping channel, while the $\Gamma_{\text{CDW}}$
keeps almost constant {[}Figs. \ref{4}(a)-(c){]}. In this process,
the plasmon energy slightly increases, but more importantly, the plasmon
loss feature becomes much more well defined with narrower linewidth
compared to the loss feature at high temperature. Besides, the plasmon
dispersion becomes close to zero even negative {[}Fig. \ref{4}(b){]}.
More simulation data with the continuous evolution can be seen in
the supplemental video \citep{SM}.

The drastic changes in the interband transition damping channel responded
to the CDW gap evolution result in the anomalous sharping of plasmon
feature. Fundamentally, this dramatic response is due to the intrinsic
small-gap semiconductor (or semimetal) band structure of the normal
phases in 1\emph{T-}TiSe$_{2}$, where the interband transitions and
the plasmon are close in the energy scale and interfere with each
other sensitively \citep{Jensen1991}. This condition is often not
satisfied in regular metal or large-gap semiconductors. This work
presents a general framework to understand how the collective electron
excitations respond to electron-hole pair excitations in semimetals
and small-gap semiconductors and provides possible tunability of plasmons
in plasmonic applications \citep{Zhu2016,Taliercio2019,Cong2020}.
\begin{acknowledgments}
The authors would like to acknowledge Prof. E. W. Plummer for his
advises in the initial stage of this study. This work was supported
by the National Key R\&D Program of China (No. 2021YFA1400200 and
No. 2017YFA0303600), the National Natural Science Foundation of China
(No. 11874404 and No. 11974399), and the Strategic Priority Research
Program of Chinese Academy of Sciences (No. XDB33000000). V. M. S.
acknowledges support from the Spanish Ministry of Science and Innovation
(Grant No. PID2019-{}-105488GB-{}-I00). X. Z. was partially supported
by the Youth Innovation Promotion Association of Chinese Academy of
Sciences.
\end{acknowledgments}

\bibliographystyle{apsrev4-2}
\phantomsection\addcontentsline{toc}{section}{\refname}\bibliography{5D__DATA_TiSe2_TiSe2_PRL_reply_2_TiSe2_PRL_DefinitiveEdition_TiSe2_Bibtex}

\end{document}